\documentclass[useAMS,usenatbib]{mn2e}
\usepackage{amsmath,amssymb}
\usepackage{natbib}
\usepackage{graphicx}
\usepackage{times}

\def\ifnote{\iffalse}

\title[GRB spectrum: a clue]
{The spectrum of Gamma-ray Burst: a clue}
\author[Y. Z. Fan]{Yi-Zhong Fan$^{1,2,3}$  \thanks{Email: Yizhong@nbi.dk}
\\
$^{1}${Niels Bohr International Academy, Niels Bohr Institute,
University of Copenhagen, Blegdamsvej 17, DK-2100
Copenhagen, Denmark} \\
$^2${\sl Dark Cosmology Centre, Niels Bohr Institute, University of Copenhagen, Juliane Maries Vej 30, DK-2100, Copenhagen, Denmark}\\
$^{3}${Purple Mountain Observatory, Chinese Academy of Sciences,
Nanjing 210008, China}
}

\begin{document}
\date{\today}
\maketitle
\label{firstpage}

\begin{abstract}
In this work we numerically calculate the thermal radiation
efficiency of the baryonic outflow. The possible outflow
acceleration in the transparent stage, which
lowers thermal radiation efficiency, has been taken into account.
In the standard internal shock model for the prompt emission, the fast shells should move with a typical Lorentz factor $\gtrsim 5 \Gamma_{\rm i}$ otherwise the GRB efficiency will be in disagreement with the observations, where $\Gamma_{\rm i}$ is the bulk Lorentz factor of the shocked/emitting region. The photosphere radius of these fast shells is small and the thermal radiation is too strong to be effectively outshone by the internal shock emission. This is particularly the case for some extremely bright events having $\Gamma_{\rm i} \sim 10^{3}$, like GRBs 080319B and 080916C. The absence of a distinct thermal component in the spectrum of most GRBs challenges the standard internal shock model and may suggest a non-baryonic (magnetic) outflow component. Though the magnetic outflow model seems favored by more and more data, it can hardly reproduce the typical GRB spectrum. In the photosphere-gradual magnetic dissipation scenario, the spectrum cuts off at $\sim 1$ GeV, too low to account for the observations of GRBs 080916C. In the sudden magnetic energy dissipation model, the low energy spectrum is expected to be $F_\nu \propto \nu^{-1/2}$, too soft to be consistent with the data $F_\nu \propto \nu^{0}$. We speculate that the low energy spectrum puzzle could be unveiled by the mechanism that particles, in the magnetic dissipation process, are repeatedly accelerated.
\end{abstract}


\section{Introduction}\label{sec:intr}
The mechanism that produces the prompt $\gamma$-ray emission in
gamma-ray burst (GRBs) is still unclear. So is the physical composition
of the GRB outflows. In the standard internal shock scenario,
the prompt soft $\gamma-$rays are the synchrotron radiation of
the shock heated electrons and the outflows are baryonic. However, there is
an increasing interest in the magnetic fireball model, in which a considerable fraction
of the outflow energy is in the form of magnetic field \citep[e.g.,][]{usov92,dt92,tho94,LB03}. Quite
a few pieces of independent evidences suggest that the GRB central engine might be strongly magnetized.
First, the analysis of some well-studied optical flashes of GRBs reveal that the magnetic fields in
 the reverse-shock region are much stronger than that in the forward-shock region, so that
  the GRB outflows are probably magnetized \citep{Fan02,ZKM03,KP03,Gomboc09}. Second, the absence of a distinct
  thermal spectrum component in most GRBs is consistent with the Poynting-flux dominated outflow model
 \citep{DM02,LB03}. Third, the non-detection of bright optical flash in most GRB afterglows can be attributed to
a mild or high magnetization of the outflow
\citep{Fan04,ZK05,Mimica09}. Fourth, the absence of a GeV-TeV
spectrum excess in the prompt emission of most GRBs detected by Fermi satellite is in agreement with the magnetic fireball model \citep{Fan09}.
Last, the (possible) detection of the high linear polarization degree of some GRBs suggests that the magnetic field
involved in the synchrotron radiation could be globally ordered \citep{Lyu03,Granot03,Gotz09}.

Very recently, two discoveries rendered the magnetic fireball model more attractive.  One is the detection and the
successful optical polarimetry of the optical flash of GRB 090102. Its optical afterglow emission declined as $\sim t^{-1.6}$ and then got shallowed to $\sim t^{-0.9}$ \citep{Gendre09}. Such behaviors can be interpreted as
the weakly magnetized reverse-shock emission superposed with the forward-shock emission. If correct,
these optical flash photons would be moderately or even highly polarized. The ongoing
polarization analysis seems to confirm such a speculation \citep{Mund09}. The data, however, has not been released yet, hampering us to go further. The other is the detection
of the featureless Band-type spectrum of GRB 080916C in a very wide energy range
$8~{\rm keV}-13~{\rm GeV}$ \citep{Abdo09}. If the GeV photons and the soft $\gamma$-rays
were from the same region, the bulk Lorentz factor of the emitting region $\Gamma_{\rm i}$ has to be
in order of $10^{3}$ and the emitting radius is $R_{\rm \gamma}\sim 2\Gamma^2 c \delta t \sim 6\times 10^{15}~{\rm cm}~(\Gamma_{\rm i}/10^{3})^{2}(\delta t/0.1~{\rm s})$, where the typical variability timescale $\delta t$ is measured in the local frame of the burst \citep{Abdo09,zfp09} and $c$ is the speed of light. For such a large $R_{\rm \gamma}$, the widely discussed photosphere
model of GRBs \citep{tho94,rm05,peer06} failed. The absence of a thermal component in the low energy band has been taken as a piece of evidence for the Poynting-flux dominated outflow model \citep{ZhangP09}, in which
the initial radius of the outflow getting accelerated is taken as $R_0 \sim c \delta t \sim 10^{9}$ cm. In principle,
$R_0$ can be as small as $\sim 10^{6}$ cm, the physical size of a stellar black hole or a magnetar\footnote{If the rate of the accretion onto the nascent black hole is high up to $\geq 1~M_\odot~{\rm s^{-1}}$, the accretion disk flow becomes optically thick to neutrinos inside a radius $\sim 10R_{\rm s}$, where $R_{\rm s}=8.85\times 10^{5}~{\rm cm}~(M_{\rm BH}/3M_\odot)$ is the black hole Schwarzchild radius and $M_{\rm BH}$ is the mass of the black hole \citep{Di02}. Most of the neutrino emission comes from outside this region. In such a case, we have $R_0 \sim 10R_{\rm s}$. On the other hand, the cooling of the disk material is dominated by neutrino radiation process, crucial for launching a baryonic outflow, only inside a radius $\sim 10^{8}$ cm \citep{Narayan01}. Therefore, in our calculation $R_0$ ranges from $10^{6}$ cm to $10^{8}$ cm, which is also consistent with what people find in the GRB spectrum modeling \citep[e.g.,][]{Ryde06,peer08,Gao09}.}. As revealed in \citet{Nakar05}, the thermal radiation from a baryonic outflow depends on $R_0$ sensitively. A small $R_0$ can suppress the thermal emission effectively. With a similar argument, \citet{Toma09b} suggested that a non-baryonic outflow component was not needed if $R_0 \sim 10^{6}$ cm.
In this work, we re-address that problem.
We show that the thermal radiation efficiency of a baryonic outflow does increase with $R_0$ rapidly, in agreement with \citet{Nakar05} and \citet{Toma09b}. However, for GRB 080916C, as long as $\Gamma_{\rm i}\sim 10^{3}$, the standard internal shock model is hard to reproduce the data even for a $R_0$ as small as $\sim 10^{6}$ cm. The physical reason is the following (see section 2 for the detailed numerical approach). In the internal shock model, the fast shells carrying most of the energy should move with a bulk Lorentz factor $\Gamma_{\rm f}\sim 2 \Gamma_{\rm sh}\Gamma_{\rm i} \gg \Gamma_{\rm i}$, where
$\Gamma_{\rm sh}$ is the strength of the internal shocks (please see section \ref{sec:IS-strength} for the discussion). The photosphere radius \citep{Pacz90,DM02,Nakar05} should be $R_{\rm ph} \sim 6\times 10^{20}~{\rm cm}~(L/10^{54}~{\rm erg~s^{-1}})\Gamma_{\rm f}^{-3}$, much smaller than $R_{\rm ph} \sim 6\times 10^{20}~{\rm cm}~(L/10^{54}~{\rm erg~s^{-1}})\Gamma_{\rm i}^{-3}$ that has been adopted in previous estimates, where $L$ is the total luminosity of the
baryonic outflow. It is well known that $\Gamma_{\rm f} \leq R_{\rm ph}/R_{0}$ \citep{Piran93,Mesz93}, so we have \begin{equation}
\Gamma_{\rm f} \leq 5\times 10^{3}L_{54}^{1/4}R_{0,6}^{-1/4}.\label{eq:Gamma_f}
\end{equation}
Please note that the convention
$Q_x=Q/10^x$ has been adopted in cgs units. Hence $\Gamma_{\rm sh}\approx \Gamma_{\rm f}/2\Gamma_{\rm i} \leq
2.5L_{54}^{1/4}R_{0,6}^{-1/4} \Gamma_{\rm i,3}^{-1}$, which is only marginally consistent with the GRB efficiency request (see footnote \ref{footnote:1} for the details).

This work is structured as the following. In section 2 we numerically calculate the thermal radiation efficiency of the baryonic outflow. We then discuss the implication of the observation of GRB 080916C on the physical composition of its outflow. In section 3 we discuss the spectrum problem in the magnetic fireball model and speculate about a possible solution. We summarize our results with some discussion in section 4.

\section{Thermal radiation expected in standard internal shock model vs. the data: shedding light on the physical composition of the GRB outflow}
The thermal radiation from the GRB outflow has been widely discussed \citep{Pacz90,tho94,mr00,DM02,rm05,Nakar05,gian06,peer06,peer08,ioka07,ZhangP09}.
In this work we focus on the thermal radiation efficiency.
Our approach is as follows. We numerically solve the number, momentum and energy conservation laws
of an extremely hot shell and get the evolution of the bulk Lorentz factor $\Gamma$, the comoving thermal
energy density $e'$, the comoving number density $n'$ and the observed surface temperature $T_{\rm obs}=\Gamma T'$
($T'$ is the comoving surface temperature of the shell). The calculation
stops when the radius reaches
\[
R_{\rm ph}\approx 6\times 10^{18}~{\rm cm}~L_{52}\eta^{-1}\Gamma^{-2},
\]
at which the thermal photons
escape from the shell, where $\eta=L/\dot{M}c^{2}$ is the dimensionless entropy and $\dot{M}$ is the mass loading rate. For $R>R_{\rm ph}$ (i.e., the transparent stage), the shell may still be accelerated by radiation via photon drag \citep{Mesz93}. Following
\citet{Rossi06}, the acceleration of the outflow by the radiation in the transparent stage can be estimated as
\begin{equation}
{d\Gamma \over dR}\approx {\sigma_{\rm T} L_{\rm ph}\over 16\pi R^2 m_{\rm p}c^{3}\Gamma^{2}}(1-{R_{\rm ph}^{4} \over R^{4}}),
\end{equation}
where the subscript ${\rm ph}$ represents the parameter at the photosphere radius, $m_{\rm p}$ is the rest mass of the proton and $\sigma_{\rm T}$ is the Thompson cross section. The final bulk Lorentz factor of the outflow is then given by
\begin{equation}
\Gamma_{\rm final}^{3}\approx \Gamma_{\rm ph}^{3}+{6L_{\rm ph} \over 5 L}\Gamma_{\rm ph}^{2} \eta.
\end{equation}
For $L_{\rm th}/L \approx e'_{\rm ph}/(e'_{\rm ph}+n'_{\rm ph}m_{\rm p}c^{2}) \ll 1$ and then $\Gamma_{\rm ph} \sim \eta$, we have $\Gamma_{\rm final} \approx \Gamma_{\rm ph}[1+0.4 \eta e'_{\rm ph}/\Gamma_{\rm ph}(e'_{\rm ph}+n'_{\rm ph}m_{\rm p}c^{2})]$. Hence the
thermal radiation efficiency can be estimated by (usually the thermal radiation from the shell surface at $R<R_{\rm ph}$ is ignorable)
\begin{equation}
\eta_{\rm th} \approx (1-0.4 {\eta \over \Gamma_{\rm ph}}){e'_{\rm th}\over e'_{\rm th}+n'_{\rm th}m_{\rm p}c^{2}}.
\label{eq:eta-th}
\end{equation}

In the case of $L_{\rm th}/L\sim 1$ (i.e., $\eta \gg \Gamma_{\rm ph} \sim 5\times 10^{3}L_{54}^{1/4}R_{0,6}^{-1/4}$), we have $\Gamma_{\rm final} \approx \Gamma_{\rm ph}(1+1.2\eta/\Gamma_{\rm ph})^{1/3}\approx 1.1 \Gamma_{\rm ph}^{2/3}\eta^{1/3}\ll \eta$ and then
$\eta_{\rm th} \sim 100\%$.

\subsection{The internal shock strength expected in typical GRBs}\label{sec:IS-strength}
For our purpose, the typical physical parameters of GRBs, in particular
the peak of the $\nu F_\nu$ spectrum $\varepsilon_{\rm p}$, the
$\gamma-$ray luminosity ($L_\gamma$), the bulk Lorentz factor of the emitting region
$\Gamma_{\rm i}$ and the emitting radius $R_\gamma$, are needed.
For the bright GRBs detected by BATSE, the distribution of $\varepsilon_{\rm p}$ peaks at
$\sim 200$ keV \citep{preece00}.  For the bursts with a known redshift $z$ detected so far, the averaged redshift is
about 2. So in the burst frame, the typical peak energy should be $(1+z)\varepsilon_{\rm p}\sim
600$ keV. As shown in \citet{Li08}, the distribution of $L_\gamma$ for 64 long-duration Swift
GRBs peaks at $\sim 5\times 10^{51}~{\rm erg~s^{-1}}$. The bulk Lorentz factor has been
derived in various ways \citep[e.g.,][]{ls01,Molinari07,xue09,zp09} and $\Gamma_{\rm i} \sim 300$
seems quite reasonable. $R_\gamma$ can be estimated by $\sim 2\Gamma_{\rm i}^{2} c \delta t$. In this work we
adopt the intrinsic variability timescale of the prompt emission $\delta t\sim 0.05$ s, as suggested by the numerical simulation of the collapsar \citep{mw99}.
So $R_\gamma$ is expected to be in order of $10^{14}$ cm. Larger $R_\gamma$ is possible, as found in
some previous estimates \citep{Zhang06,LB06,kumar07}.

Following Fan \& Piran (2008), the magnetic field strength $B$ and the
typical random Lorentz factor $\gamma_{\rm m}$ can be estimated as the following.
The $\gamma-$ray luminosity $L_\gamma$ is related to the total
luminosity of the emitting material $L$ as $L_\gamma \sim
\epsilon_{\rm e}L/(1+\bar{Y})$, where $\epsilon_{\rm e}$ ($\epsilon_{\rm B}$) is the fraction of the shock energy given to the electrons (magnetic field) and $\bar{Y}$ is the averaged Compton parameter.
The comoving strength of the magnetic field can thus be
estimated by
\begin{eqnarray}
B &\approx & (2\epsilon_{\rm B}L/R_\gamma^2 \Gamma_{\rm i}^2
c)^{1/2}\nonumber\\
&\approx& 4.5\times 10^{4}~{\rm Gauss}~({\epsilon_{\rm B}\over
\epsilon_{\rm e}})^{1\over 2}(1+\bar{Y})^{1\over 2}L_{\rm
\gamma,52}^{1\over 2}R_{\gamma,14}^{-1}\Gamma_{\rm i,2.5}^{-1}.
\end{eqnarray}

The synchrotron radiation frequency of
electrons with a typical Lorentz factor $\gamma_{\rm m}$ is
$\sim (1+z)\varepsilon_{\rm p}/h$, which in turn yields
\begin{eqnarray}
\gamma_{\rm m} &\sim & [{(1+z)\varepsilon_{\rm p} \over 2.8\times
10^{6} h \Gamma_{\rm i} B }]^{1/2}\nonumber\\
&\sim & 2000~[{(1+z)\varepsilon_{\rm p}\over 600~{\rm keV}}]^{1\over
2}({\epsilon_{\rm e}\over \epsilon_{\rm B}})^{1\over
4}[(1+\bar{Y})L_{\gamma,52}]^{-1\over 4}R_{\gamma,14}^{1\over 2},
\label{eq:gamma_em1}
\end{eqnarray}
and then
\begin{eqnarray}
\Gamma_{\rm sh}&\sim & 1+1\zeta_{\rm e,-1}[{(1+z)\varepsilon_{\rm p}\over 600~{\rm keV}}]^{1\over
2}({\epsilon_{\rm e}\over \epsilon_{\rm B}})^{1\over
4}[(1+\bar{Y})L_{\gamma,52}]^{-1\over 4}R_{\gamma,14}^{1\over 2}\nonumber\\
 &&[(p-1)/3(p-2)](\epsilon_{\rm e}/0.3)^{-1},\label{eq:Gamma_sh-I}
\end{eqnarray}
where $h$ is the Planck's constant,
 $0<\zeta_{\rm e}\leq 1$ is the fraction of the electrons getting accelerated at the shock front, and $p$ is the index of the power-law energy distribution of the accelerated electrons.

One can see that even for $\zeta_{\rm e} \sim 0.1$, the shocks are relativistic, i.e.,
$\Gamma_{\rm sh} \sim 2$. Such an estimate is likely conservative since
a larger $\zeta_{\rm e} \sim 1$ is needed to account for the early X-ray afterglow observations
\citep[][section 3.2 therein]{FP06}. Please note that $\Gamma_{\rm sh}\geq 2$ is also required to
get a GRB efficiency\footnote{Let's just consider the collision of two shells. The masses and the Lorentz factors of the fast and slow shells are denoted as $(M_{\rm f},~M_{\rm s})$ and $(\Gamma_{\rm f},~\Gamma_{\rm s})$, respectively.  The internal shocks are most efficient when an inner engine produces shells with comparable energy but very different Lorentz factors, i.e., $\Gamma_{\rm f}M_{\rm f}=\Gamma_{\rm s}M_{\rm s}$ \citep{KPS97}.
In such a case the merged shell has a bulk Lorentz factor $\Gamma_{\rm i}\sim \sqrt{2}\Gamma_{\rm s}$ and the efficiency is $\eta_{\rm i}\approx 1-(M_{\rm f}+M_{\rm s})\Gamma_{\rm i}/(\Gamma_{\rm f}M_{\rm f}+\Gamma_{\rm s}M_{\rm s})\approx 1-(M_{\rm f}/M_{\rm s}+1)/\sqrt{2}$ \citep{Piran99}. Setting $\eta_{\rm i}\sim 20\%$, we have $M_{\rm f}/M_{\rm s} \sim 0.14$, $\Gamma_{\rm f}\sim 7\Gamma_{\rm s} \sim 5\Gamma_{\rm i}$ and $\Gamma_{\rm sh}\sim 2.5$. \label{footnote:1}} $\eta_{\rm i} \sim 20\%$. Therefore, in the internal shock model, for typical bright GRBs, we have $\Gamma_{\rm f} \sim 2\Gamma_{\rm sh}\Gamma_{\rm i} \sim 1000 \Gamma_{\rm i,2.5}$. For extremely bright
bursts, like GRB 080319B and GRB 080916C, $\Gamma_{\rm f} \sim 5\times 10^{3}$ is needed since
$\Gamma_{\rm i}\sim 10^{3}$ and $R_\gamma > 10^{15}$ cm \citep{zfp09,Abdo09}.

\subsection{The thermal radiation leaking from the surface at $R<  R_{\rm ph}$}\label{sec:thermal-inner}
The acceleration of one baryonic shell with a
width $c\delta t$ (measured by
the observer) is driven by the thermal photons and can be
approximated as $\Gamma \sim R/R_0$ for $R<R_*\equiv \eta R_0<R_{\rm ph}$
\citep{Piran93,Mesz93}. The thermal emission from the
surface of a shell in the case of $R<R_*(<R_{\rm ph})$ can be estimated as
\begin{equation}
L_{\rm th,s}\sim 4\pi R^2 \sigma {T'}^4 \Gamma^2,
\end{equation}
so the total energy emitted during the acceleration phase is
\begin{equation}
E_{\rm th,s} \sim \int L_{\rm th,s} dR/(\Gamma^2 c) \sim 4\pi \sigma R_0^2 T_0^4
(R_0/c),
\end{equation}
where the relation $T_{\rm obs} \sim $ const. \citep{Piran93,Mesz93} has been taken into account, $T_0$ is the
temperature of the initial outflow and $\sigma$ is the Stefan-Boltzmann constant.

The total energy of the shell can be estimated as
\begin{equation}
E_{\rm tot} \sim 4\pi R_0^2 c a T_0^4\delta t,
\end{equation}
where $a\equiv 4\sigma/c$ is the radiation constant.

The GRB efficiency contributed by the thermal
emission leaking from the surface at $R\leq R_*$ is then given by
\begin{equation}
\eta_{\rm th,s} \sim {E_{\rm th,s} \over E_{\rm tot}} \sim {R_0 \over 4c\delta t}.
\end{equation}
Usually the central engine has a radius $R_0 \geq 10^{6}$ cm. The typical
variability timescale of the GRB outflow may be mainly governed by the
accretion process and can be as long as $\sim 50$ ms \citep{mw99}. For these
typical parameters, we find
\begin{equation}
\eta_{\rm th,s} \sim 1.5\times 10^{-4}~R_{0,6}(\delta t/50~{\rm ms})^{-1}.
\end{equation}
Therefore the thermal radiation of the accelerating shell is unimportant unless $\delta t\sim R_0/c$
or $R_{\rm ph}<R_*$.

For $R_*<R<R_{\rm ph}$, $T'\approx \eta^{-1} T_0 (R/R_*)^{-2/3}$ \citep{Piran93,Mesz93},
the subsequent thermal emission should have a luminosity $L_{\rm th,s-l}\approx 4\pi \eta^2 R^2 \sigma {T'}^{4}\approx
4\pi R_0^{2}\sigma T_0^{4}(R/R_*)^{-2/3}$ and the detected
emission should have a total energy $E_{\rm th,s-l}=\int^{R_{\rm ph}}_{R_*}4\pi R_0^{2}\sigma T_0^{4}(R/R_*)^{-2/3}dR/(2\eta^2 c)=3(R_{\rm ph}/R_*)^{1/3}R_*(4\pi R_0^{2}\sigma T_0^{4})/(2\eta^{2} c) $.
Since $(R_{\rm ph}/R_*)^{1/3} \sim 10 \ll \eta$, we have $E_{\rm th,s-l} \ll E_{\rm th,s}$.

\subsection{The thermal radiation efficiency of the baryonic outflow}\label{sec:thermal}
Following \citet{Piran93} and \citet{KG02}, for an extremely hot outflow
we have the following number, energy and momentum conservation laws
\begin{equation}
{1\over c}{\partial (n'\Gamma)\over \partial t}+{1\over R^2}{\partial (R^{2}n'u)\over \partial R}=0,
\label{eq:basic1}
\end{equation}
\begin{equation}
{1\over c}{\partial (w'\Gamma^{2})\over \partial t}+{1\over R^2}{\partial [R^{2}w' \Gamma u]\over \partial R}={1\over c}{\partial p' \over \partial t}-{\Gamma \Lambda'\over c},
\label{eq:basic2}
\end{equation}
\begin{equation}
{1\over c}{\partial (w' \Gamma u)\over \partial t}+{1\over R^2}{\partial (R^{2}w' u^{2})\over \partial R}=-{\partial p' \over \partial R}-{u \Lambda' \over c},
\label{eq:basic3}
\end{equation}
where the comoving entropy density $w'=n'm_{\rm p}c^{2}+\hat{\gamma}e'$, the thermal pressure denisty $p'=(\hat{\gamma}-1)e'$, $u=\Gamma \beta=\sqrt{\Gamma^{2}-1}$, and $\Lambda' \sim \sigma {T'}^4/c \Gamma \delta t \sim e'/(4\Gamma \delta t)$. The specific heat ratio can be estimated by $\hat{\gamma}\approx 1+(a{T'}^{4}/3+n' k {T'})/(a{T'}^{4}+3n' kT'/2)$. In the current case we find that $a{T'}^{4}/3\gg n' k T'$, for which $\hat{\gamma}\approx 4/3$.

In the case of $\Gamma\gg 1$, the above equations can be significantly simplified.
Elimination of the radiative cooling term from equations (\ref{eq:basic2}) and (\ref{eq:basic3}) leads to
\citep[see also][]{KG02}
\begin{equation}
{d p' \over dt}+\Gamma^{2}\beta w'{d \beta \over dt}\approx 0,
\label{eq:new1}
\end{equation}
where the convective derivative is $d/dt=\partial/\partial t + \beta c \partial/\partial R$.

Eqs.(\ref{eq:basic1}) and (\ref{eq:basic3}) can also be approximated as
\begin{equation}
{1\over c}{d (\Gamma n') \over dt}+{2 \Gamma n' \over R} \approx 0
\label{eq:new2}
\end{equation}
and
\begin{equation}
{1\over c}{d (\Gamma^{2} \beta w') \over dt}+{2 \Gamma^{2} \beta w' \over R} \approx -{u \Lambda' \over c},
\label{eq:new3}
\end{equation}
respectively.

Eqs.(\ref{eq:new1}-\ref{eq:new3}), together with the relation $dR=\beta c dt$, are complete for solving $\Gamma$, $e'$ and $n'$ as functions of $R$. The starting point in our calculation is $R=5R_0$, at which we take
$\Gamma=5$. The calculation ends at $R=R_{\rm ph}$.

\begin{figure}
\begin{picture}(0,180)
\put(0,0){\includegraphics{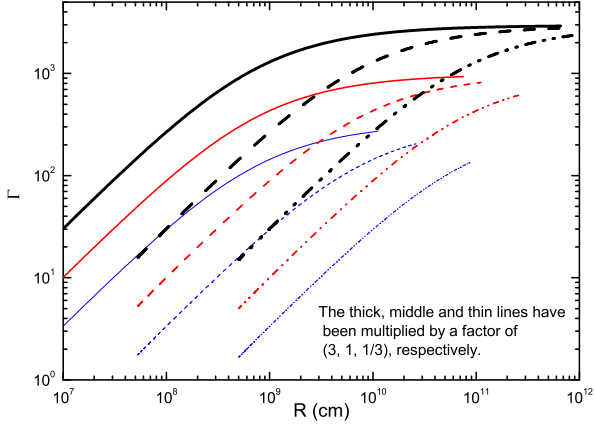}}
\end{picture}
\caption{The bulk Lorentz factor of the shell as a function of $R$. The solid, dashed and dash dot-dotted lines are for
$R_0=(10^{6},~10^{7},~10^{8})$, respectively. The thick, middle and thin lines are for $L=(10^{54}, ~10^{53},~10^{52})~{\rm erg/s}$, respectively. $\eta=1000$ is assumed in the calculation.} \label{fig:Gamma}
\end{figure}

\begin{figure}
\begin{picture}(0,180)
\put(0,0){\includegraphics{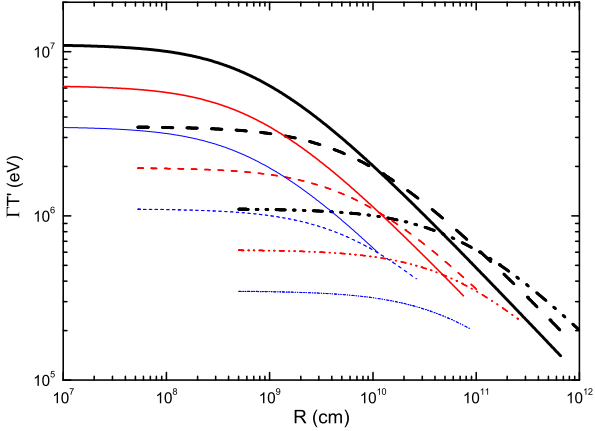}}
\end{picture}
\caption{The observed surface temperature of the shell as a function of $R$. The line styles and $\eta$ are
the same as those of Fig.\ref{fig:Gamma}.} \label{fig:R}
\end{figure}

\begin{figure}
\begin{picture}(0,180)
\put(0,0){\includegraphics{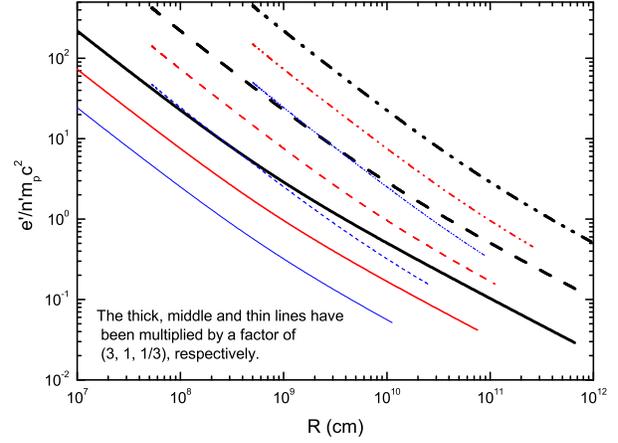}}
\end{picture}
\caption{The parameter $e'/(n'm_{\rm p}c^{2})$ of the shell as a function of $R$.
The line styles and $\eta$ are
the same as those of Fig.\ref{fig:Gamma}.
One can see that the larger the $R_0$, the higher the fraction of thermal energy escaping from the
shell at $R_{\rm ph}$.} \label{fig:e-n}
\end{figure}

Our numerical results have been plotted in Fig.\ref{fig:Gamma}-Fig.\ref{fig:e-n}. We find that $\Gamma \propto R$,
 $T_{\rm obs} \sim {\rm const.}$ and $e'/n'm_{\rm p}c^2 \propto R^{-1}$ for $e'\geq n'm_{\rm p}c^{2}$, while $\Gamma \sim {\rm const.}$, $T_{\rm obs}\propto R^{-2/3}$ and $e'/n'm_{\rm p}c^2 \propto R^{-2/3}$ for $e'<  n'm_{\rm p}c^{2}$.
All are consistent with \citet{Piran93}, as expected.
If at late times $\hat{\gamma}$ approaches a constant lying between $4/3$ and $5/3$, with the general relation $p'\propto n'^{\hat{\gamma}}$ we have
\begin{equation}
e'/n'\propto R^{-2(\hat{\gamma}-1)},~~T_{\rm obs}\propto R^{-\hat{\gamma}/2}.
\end{equation}

\begin{figure}
\begin{picture}(0,180)
\put(0,0){\includegraphics{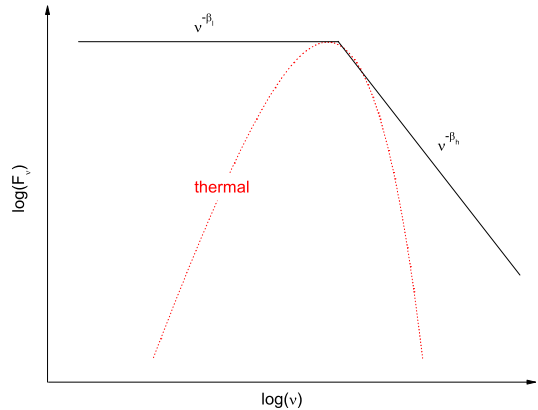}}
\end{picture}
\caption{A schematic plot of hiding the thermal emission by the nonthermal radiation powered by late energy dissipation, for example, internal shocks.} \label{fig:Cartoon}
\end{figure}

\subsection{Theoretical prediction versus the data: Constraint on the nature of the outflow}
Can the thermal emission be outshone by the nonthermal emission powered by internal shocks?
For simplicity we assume the non-thermal emission component takes the form $F_\nu =F_0 (\nu/\nu_0)^{-\beta_{\rm l}}$
for $\nu<\nu_0=\varepsilon_{\rm p}/h$ and $F_\nu=F_0 (\nu/\nu_0)^{-\beta_{\rm h}}$ for $\nu>\nu_0$. As found in the data analysis, for typical GRBs $\beta_{\rm l}\sim 0$ and $\beta_{\rm h} \sim 1.25$ \citep{preece00}. The thermal radiation peaks at a frequency $h\nu_{\rm th,p} =2.82kT_{\rm obs}$ and the corresponding flux
is $F_{\nu_{\rm th,p}}=2h\nu_{\rm th,p}^{3}/c^{2}/[\exp(h\nu_{\rm th,p}/kT_{\rm obs})-1]$. In the case of $\beta_{\rm l}<1$ and
 $\beta_{\rm h}>1$ (see Fig.\ref{fig:Cartoon} for the details), in order to hide the thermal emission component, the ratio between the thermal emission energy $E_{\rm th}$ and nonthermal emission energy $E_{\rm nth}$ should satisfy
 \begin{equation} {E_{\rm nth} \over E_{\rm th}} \geq
{0.6 (\beta_{\rm h}-\beta_{\rm l}) \over (1-\beta_{\rm l})(\beta_{\rm h}-1)}\left\{%
\begin{array}{ll}
    ({\nu_0\over \nu_{\rm th,p}})^{1-\beta_{\rm l}}, & \hbox{for $\nu_{\rm th,p}<\nu_0$;} \\
    1, & \hbox{for $\nu_{\rm th,p}=\nu_0$;} \\
    ({\nu_{\rm th,p} \over \nu_0})^{\beta_{\rm h}-1}, & \hbox{for $\nu_{\rm th,p}>\nu_0$.} \\
\end{array}%
\right.
\end{equation}
For typical GRBs, $E_{\rm nth} \geq 3-10 E_{\rm th}$ is needed, otherwise the thermal component
can not be hidden. The GRB internal shocks should have an efficiency
$\eta_{\rm i}\sim 20\%$ of converting the kinetic energy of the outflow into radiation, as found
in the afterglow modeling \citep[e.g.,][]{FP06}. As a result,
the thermal radiation efficiency should satisfy $\eta_{\rm th} \leq \eta_{\rm i}/10 \sim 2\%$.  Since the fast shells just take a fraction of the total energy \footnote{For the most efficient internal shocks suggested in \citet{KPS97},
$\Gamma_{\rm f}M_{\rm f}\sim \Gamma_{\rm s}M_{\rm s}$, so the fraction is $\sim 1/2$.}, the limit can be a bit higher. A reasonable requirement is
\begin{equation}
\eta_{\rm th} <5\%.
\label{eq:request-I}
\end{equation}
For $L\sim 10^{52}~{\rm erg~s^{-1}}$, $\eta \sim \Gamma_{\rm f} \sim 10^{3}$, and $R_0 \sim 10^{6}$ cm,
we have $e'_{\rm ph}/(e'_{\rm ph}+n'_{\rm ph}m_{\rm p}c^2) \sim 15\%$ (see Fig.\ref{fig:e-n}), $\eta/\Gamma_{\rm ph}=1.25$, and then $\eta_{\rm th} \sim 7\%$, which violates the above request (though marginally). So the absence of a distinct thermal spectrum component in most GRBs \citep{Ryde06}
may be a problem of the standard internal shock model. The same conclusion has already been drawn by \citet{DM02}.
However in their modeling
 a very small $\zeta_{\rm e}$ is needed and the resulting internal shock efficiency is very low ($\leq $ a few percent).

\begin{figure}
\begin{picture}(0,200)
\put(0,0){\includegraphics{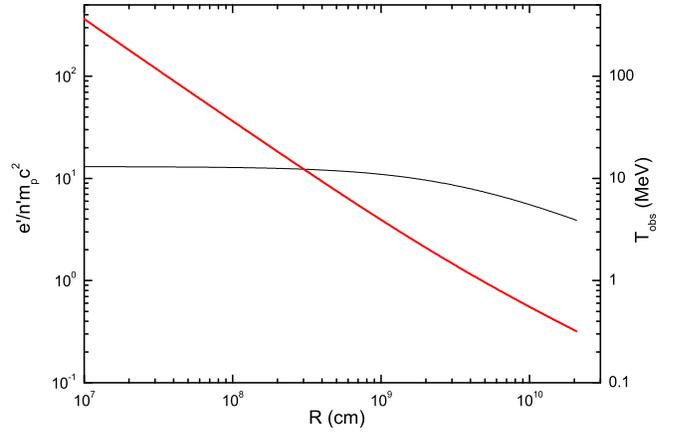}}
\end{picture}
\caption{The observed surface temperature of the shell (the thin lines) and $e'/n'm_{\rm p}c^{2}$
(the thick lines) as a function of $R$. The parameters are $\eta=5000$, $R_0=10^{6}$ cm and $L=2\times 10^{54}~{\rm erg~s^{-1}}$.} \label{fig:080916C}
\end{figure}

If {\bf GRB 080916C} indeed had
$\Gamma_{\rm f}\sim 5\times 10^{3}$, the thermal radiation might be
very strong. For this burst, the afterglow data had not been collected until
half a day after the trigger \citep{Greiner09}. The data are rare and can be well understood
within the forward shock model supposing the medium is a very weak stellar wind \citep{zfp09}. The physical
parameters can not be uniquely determined. The isotropic-equivalent kinetic energy of the
outflow can be as large as $\sim 4\times 10^{55}$ erg and the corresponding GRB efficiency is
$\eta_{\rm i} \sim 20\%$ \citep{Gao09}.
As shown in Fig.\ref{fig:080916C}, for $L\sim 2\times 10^{54}~{\rm erg~s^{-1}}$
and $R_0 \sim 10^{6}$ cm, we have $\eta_{\rm th} \sim 10\%$, violating
 eq.(\ref{eq:request-I}). A thermal spectrum component will be distinct. The absence of such a component
may thus favor the non-baryoinc (plausibly magnetic) outflow model.
\citet{ZhangP09} got the same conclusion. However in their approach
 $R_0 \sim 10^{9}~\rm cm$ is assumed, much larger than what we adopt.

\section{The spectrum problem of GRBs in the magnetized outflow model}
The absence of a distinct thermal component
is at odds with the standard internal shock model and may favor
the magnetic fireball model. A self-consistent interpretation
of the typical Band spectrum of GRBs, however, is still unavailable, as shown below.

In a pure photosphere model
in the case of magnetar wind for GRBs, \citet{tho94} showed that the typical Band spectrum of GRBs
could be reproduced. However, for GRB 080916C, the prompt emission with a single power-law spectrum up to
$\sim 70 (1+z)^{-1}$ GeV suggests that $R_\gamma \sim 10^{16}$ cm, much larger than the site
of the photosphere $\sim 10^{9}$ cm suggested by \citet{tho94}.

\citet{gian07} showed that within the photosphere-gradual magnetic dissipation scenario,
the low energy spectrum could be as hard as $F_\nu \propto \nu^0$. However, the corresponding high energy spectrum
was usually a bit harder than $F_\nu \propto \nu^{-1}$, inconsistent the data. What's worse,
\citet{gian07} predicted a cutoff at an energy $\sim 1$ GeV, too low to account for
the observations of GRB 080916C.

Below we focus on the model of a {\it sudden magnetic energy dissipation} at $R_\gamma \sim 10^{15}-10^{16}$ cm
\citep[e.g.,][]{LB03}. We introduce the parameter $\bar{\sigma}$ to denote the ratio between the magnetic and the particle energy density. We perform a general study in which the details of the
magnetic dissipation and the subsequent
particle acceleration have been ignored. After the dissipation,
the strength of the residual magnetic field can be estimated as
\begin{equation}
B_{\rm m} \sim 300~{\rm Gauss}~k_{-1}^{1/2}[\bar{\sigma}/(1+\bar{\sigma})]^{1/2}L_{\rm m,52}^{1/2}R_{\gamma,15.5}^{-1}
\Gamma_{\rm i, 2.5}^{-1}
\end{equation}
where $0\leq k\leq 1$ is the parameter reflecting the importance of the magnetic dissipation,  which has
been normalized to $0.1$ because a residual magnetization $\bar{\sigma}_{\rm d} \sim k \bar{\sigma}/[1+(1-\epsilon_{\rm e})(1-k)\bar{\sigma}]\geq 0.1$ may be needed to account for the absence of
bright optical flashes in most GRB afterglows. {\it Obviously $k$ plays the same role of $\epsilon_{\rm B}$
in estimating the strength of magnetic field of the emitting region.}
The electrons accelerated by the energy dissipation are assumed to take a power-law distribution $\propto \gamma_{\rm e}^{-p}$ for $\gamma_{\rm e}>\gamma_{\rm m}$. Similar to eq.(\ref{eq:gamma_em1}) we have
\begin{eqnarray}
\gamma_{\rm m} &\sim & 2.5\times 10^{4}~[(1+z) \varepsilon_{\rm p}/600~{\rm keV}]^{1/2}k_{-1}^{-1/4}L_{\rm 52}^{-1/4}\nonumber\\
&&[\bar{\sigma}/(1+\bar{\sigma})]^{-1/4}R_{\gamma,15.5}^{1/2}.
\label{eq:gamma_em3}
\end{eqnarray}
The corresponding constraint on $\bar{\sigma}$ reads
\begin{eqnarray}
\bar{\sigma} &\sim & 140~\zeta_{\rm e} [3(p-2)/(p-1)]^{-1}[(1+z) \varepsilon_{\rm p}/600~{\rm keV}]^{1/2}\nonumber\\
&&(\epsilon_{\rm e}/0.3)^{-1} k_{-1}^{-1/4}L_{\rm 52}^{-1/4}
[\bar{\sigma}/(1+\bar{\sigma})]^{-1/4}R_{\gamma,15.5}^{1/2}.
\label{eq:sigma}
\end{eqnarray}
The outflow has to be highly magnetized otherwise the dissipated energy is not enough to
accelerate electrons to a typical random Lorentz factor $\sim {\rm a~few\times} 10^{4}$.

The cooling Lorentz factor can be estimated as
\begin{equation}
\gamma_{\rm c} \sim 50~L_{\rm 52}^{-1}[\bar{\sigma}/(1+\bar{\sigma})]^{-1}k_{-1}^{-1}R_{\gamma,15.5}\Gamma_{\rm i, 2.5}^{3}.
\end{equation}
We need a very large
\begin{equation}
\Gamma_{\rm i} \sim 5000~L_{\rm 52}^{1/4}[\bar{\sigma}/(1+\bar{\sigma})]^{1/4}k_{-1}^{1/4}R_{\gamma,15.5}^{-1/6}[(1+z) \varepsilon_{\rm p}/600~{\rm keV}]^{1/6}.
\label{eq:Gamma}
\end{equation}
to get $\gamma_{\rm c} \sim \gamma_{\rm m}$ and then a low energy spectrum
$F_\nu \propto \nu^{1/3}$ that is roughly consistent with the observations.
Such a large $\Gamma_{\rm i}$ is unrealistic (see eq.(\ref{eq:Gamma_f}) for the constraint)
and is in contradiction with other constraints. So the low energy spectrum is likely $F_\nu \propto \nu^{-1/2}$,
inconsistent with the data. Such an inconsistence between the model and the data, already found in standard internal shock model, is the
 so-called ``low energy spectrum crisis of GRBs" \citep[e.g.,][]{Cohen97,ghis00,km08,psz09}.

Eqs.(\ref{eq:sigma}) and (\ref{eq:Gamma}) suggest a very low baryon loading of the outflow
\[M_{\rm jet} \sim E_{\rm jet}/(\Gamma \bar{\sigma} c^{2}) \sim 5\times 10^{-10}~M_\odot~\zeta_{\rm e}^{-1} \Gamma_{\rm i,2.5}^{-1}E_{\rm jet,51},\]
where $E_{\rm jet}$ is the typical geometry-corrected energy of GRBs.
It is unclear how such clean fireballs
can be launched in the collapsar scenario. A hot massive neutron star as the
GRB central engine is disfavored because of the huge baryon pollution from such a star \citep[e.g.,][]{LE93}.\\

\begin{figure}
\begin{picture}(0,180)
\put(0,0){\includegraphics{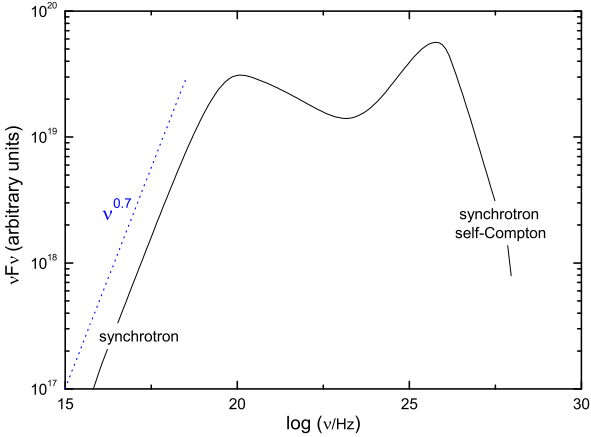}}
\end{picture}
\caption{The synchrotron and synchrotron self-Compton spectrum of the electrons suffering
significant inverse Compton cooling though within Klien-Nishina regime. The parameters are
$p= 2.5$, $\gamma_{\rm m} =7\times 10^{3}$, $\Gamma_{\rm i} =200$, $g \sim 35$ and $\epsilon_{\rm e}/k = 100$.} \label{fig:DKK}
\end{figure}

In the above estimate of the synchrotron spectrum we have not taken into account the fact that the IC cooling of electrons is energy dependent. Such a correction may be crucial if the IC scattering process takes place in the Klein-Nishina regime. With Eq.(\ref{eq:gamma_em3}) we have
$g \equiv \gamma_{\rm m}(1+z)\varepsilon_{\rm p}/\Gamma_{\rm i} m_{\rm e}c^2 \gg 1$. The factor $g$ determines the regime of
scattering of electrons with a random Lorentz factor $\gamma_{\rm m}$
on its synchrotron radiation photons at a typical energy
$\varepsilon_{\rm p}$.
The IC cooling of the electrons with $\gamma_{\rm e}>\Gamma_{\rm i} m_{\rm e}c^2/(1+z)\varepsilon_{\rm p}$ is suppressed by the Klein-Nishina effect. As a result, the synchrotron emissivity of these electrons increases with
$\gamma_{\rm e}$, which leads to a harder synchrotron spectrum and has been adopted to account for the typical GRB X-ray spectrum $F_{\nu}\propto \nu^{0}$ \citep[e.g.,][hereafter DKK]{Deri01}. However, as shown {\it in the Appendix}, the magnetic field in the emitting region should be
very low otherwise the Klein-Nishina effect is too weak to modify the energy distribution of electrons and then the
radiation spectrum (see eq.(\ref{eq:ratio})). Even for an unreasonably small $k$, say $\sim 10^{-3}-10^{-2}$, DKK's scenario may still not work. In Fig.\ref{fig:DKK} we present our numerical spectrum based on the code developed in \citet[][the instantaneous approximation with some minor modifications]{Fan08}.
The low energy spectrum
does get hardened, as widely speculated \citep[e.g.,][]{Deri01,Deri03,Deri07,wang09,nakar09,Daigne09}. However, the resulting spectrum $F_\nu \propto \nu^{-0.3}$ for $\epsilon_{\rm e}/k \sim 100$ is not hard enough
to account for the typical data $F_\nu \propto \nu^0$.
The low energy spectrum could be as hard as $F_\nu \propto \nu^{-0.1}$ if
 $\epsilon_{\rm e}/\epsilon_{\rm B}\sim 10^{4}$ \citep{nakar09}, corresponding to $k\sim 10^{-5}$. It is, however, unclear how the magnetic energy dissipation can be so efficient
 (i.e., $k \ll 1$). It is also unclear whether the resulting
 spectrum for $k\sim 10^{-5}$ can be well approximated by the
 Band function or not (see eq.(\ref{eq:ratio}) for the constraint).
Observational tests of DKK's scenario may be available soon.

A speculated solution to the low energy spectrum problem is the repeated acceleration of the particles
in the energy dissipation process but much more work is needed to see whether it is the case. Some preliminary
discussion, but only for unmagnetized outflow, can be found in \citet{sp04}. Please also bear in mind that
in the multiple particle acceleration model the magnetization is required to be (much) higher than that suggested in eq.(\ref{eq:sigma}) otherwise the re-accelerated electrons can not achieve a random Lorentz factor as large as $\sim 10^{4}$ (see eq.(\ref{eq:gamma_em3})).

\section{Conclusion and Discussion}
In this work, we have calculated the thermal radiation efficiency of a baryonic shell
(see section 2 for the details) and have discussed the spectrum problem of GRBs (see section 3 for the details). In the standard internal shock model for the prompt emission, the fast shells should move with a typical Lorentz factor $\Gamma_{\rm f} \sim 5 \Gamma_{\rm i} \sim 10^{3} \Gamma_{\rm i,2.3}$ otherwise the GRB efficiency will be in disagreement with the observations. The photosphere radius of such fast shells is $\sim 6\times 10^{9}~{\rm cm}~L_{52}\Gamma_{\rm f,3}^{-3}$, much smaller than that of the slow shells with a typical Lorentz factor $< \Gamma_{\rm i}$. Consequently the thermal radiation from fast shells will be much stronger than that from the slow shells.
In the internal shock model one should focus on the fast shells when investigating the thermal emission of GRB outflow. We find out that though most of the thermal energy has been converted into the kinetic energy of the baryons, the residual thermal photons escaping from the surface at $R\geq R_{\rm ph}$ can not be ignored. The possible acceleration of the outflow in the transparent stage via photon drag may be able to lower the thermal radiation efficiency by a factor of $\sim 0.5$. For typical GRB parameters $(L,~R_0,~\Gamma_{\rm f})\sim (10^{52}~{\rm erg~s^{-1}}~10^{6}~{\rm cm},~10^{3})$, we have a thermal radiation efficiency $\sim 7\%$  (see Tab.\ref{tab:con-2} for a summary). These thermal photons are detectable and play an important role in cooling the electrons accelerated in the internal shocks. The non-detection of such a spectrum component in most GRBs thus challenges the standard internal shock model. This is particularly the case for some extremely bright bursts with a featureless Band spectrum
and a very large $\Gamma_{\rm i}\sim 10^{3}$, like GRB 080916C. Please note that our conclusion is for $R_0 \sim 10^{6}$ cm, the lowest value it could be. For a magnetized outflow, the thermal emission is expected to be weak \citep{LB03}, consistent with the data.

\begin{table*}
\caption{The thermal radiation efficiency of fast shells involved in the standard internal shock model.}
\begin{tabular}{l|l|c|c|c|c}
\hline
 & typical $\Gamma_{\rm i}$ & corresponding $\Gamma_{\rm f}$ & $\eta_{\rm th}$   & $\eta_{\rm th}$ \\
 &   & & for $R_0=10^{6}$ cm &  for $R_0=10^{7}$ cm
\\ \hline
bright GRBs ($L_{52}=1$) & $\sim 200-300$ & $\sim 10^{3}$ &  $\sim 7\%$ & $\sim 18\%$
\\ \hline
extremely bright GRBs ($L_{54}=1$) & $\sim 10^{3}$ & $\sim 5\times 10^{3}$ & $\sim 15\%$
& $\sim 35\%$ \\
\hline
\end{tabular}
\label{tab:con-2}
\end{table*}

There is an increasing interest in the magnetic fireball model. A self-consistent interpretation
of the typical Band spectrum of GRBs, however, is still unavailable (see section 3 for the details). In the photosphere-gradual magnetic dissipation scenario, the resulting
spectrum cuts off at an energy $\sim $ a few GeV, at odds with the observations of GRBs 080916C.
In the sudden magnetic energy dissipation model, the low energy spectrum is expected to be $F_\nu \propto \nu^{-1/2}$, too soft to be consistent with the data.
It is suggested that the synchrotron radiation spectrum of electrons suffering significant
IC cooling but within Klein-Nishina regime can be much harder than the standard fast-cooling spectrum
$F_\nu \propto \nu^{-1/2}$ \citep{Deri01},
 helping us solve the so-called low energy spectrum crisis of GRBs. However, to reproduce the data, the magnetic field in the emitting region is required to be extremely low (see eq.(\ref{eq:ratio}) and the last paragraph of section 3 for the discussion), which seems unrealistic in the magnetic dissipation scenario. The particle re-acceleration may be able to give rise to a harder low energy spectrum but much more work is needed to see whether it is the case.

\section*{Acknowledgments}
This work was supported in part by the Danish National Science
Foundation, Chinese Academy of Sciences, National basic research
program of China under the grant 2009CB824800. The author
acknowledges the hospitality of Nordita at Stockholm,
where part of this work was done. I thank T. Piran, the anonymous referee, X. F. Wu and K. Toma for constructive remarks and B. Zhang, E. V. Derishev, S. Kobayashi, E. Nakar, F. Daigne, D. Giannios, K. Ioka, and Y. C. Zou for fruitful discussions and/or email communications.

\begin{appendix}
\section{DKK's scenario as a possible solution of the low energy spectrum crisis of GRBs: some requests}\label{sec:DKK}
It is widely believed that
the prompt soft $\gamma-$rays are the synchrotron radiation of
the shock heated electrons. With reasonable parameters, the synchrotron radiation
peaks in soft gamma-ray band, consistent with the data. The resulting low energy
X-ray spectrum $F_\nu \propto \nu^{-1/2}$, is, however softer than the
typical low energy spectrum $F_\nu \propto \nu^0$ \citep{preece00}.  This
inconsistency between the model and the data is sometimes called  ``low energy spectrum
crisis of GRBs" \citep[e.g.,][]{Cohen97,ghis00}.

\begin{figure}
 \includegraphics[angle=0,width=0.5\textwidth]{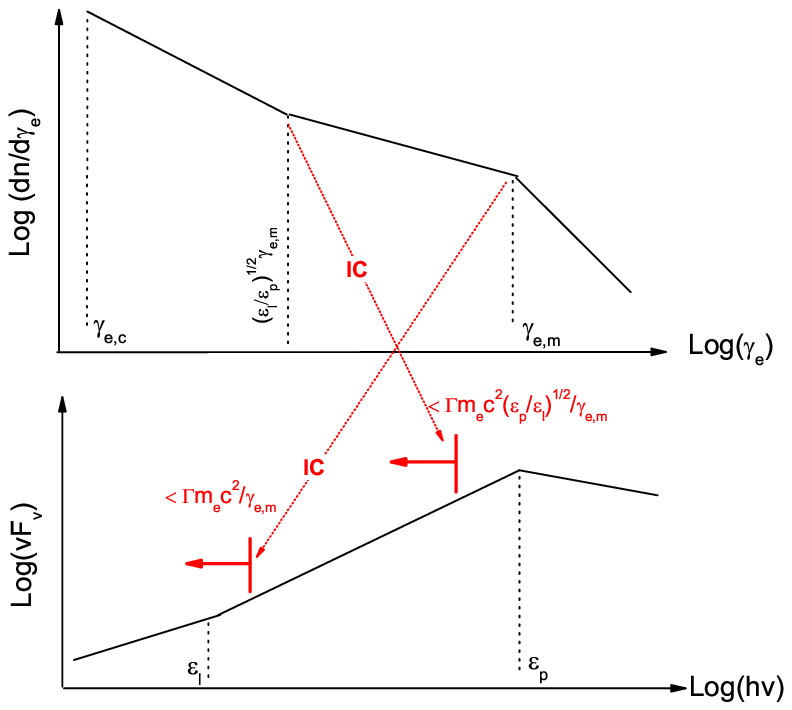}
 \caption{The schematic plot of Derishev et al. (2001)'s model.}
 \label{fig:Derishev01}
\end{figure}

In DKK's scenario (see Fig.\ref{fig:Derishev01} for illustration), to obtain a hard X-ray spectrum $F_\nu \propto \nu^0$ rather than
$\propto \nu^{-1/2}$ for $\varepsilon_{\rm l}<h\nu<\varepsilon_{\rm p}$, the emitting electrons should satisfy:
(1) $\gamma_{\rm c}<\gamma_{\rm m}(\varepsilon_{\rm p}/\varepsilon_{\rm l})^{-1/2}<\gamma_{\rm e}<\gamma_{\rm m}$.
(2) The SSC cooling of the electrons at an energy $\sim \gamma_{\rm m}m_{\rm e}c^2$ is
important (the Compton parameter $Y\geq 1$) though in the Klein-Nishina regime.
Let's show how it works. In general,
the energy distribution of fast cooling electrons can be estimated by \citep{nakar09}
\begin{equation} {dn\over d\gamma_{\rm e}} \propto [1+Y(\gamma_{\rm
e})]^{-1}
\left\{%
\begin{array}{ll}
    \gamma_{\rm e}^{-2}, & \hbox{for $\gamma_{\rm c}<\gamma_{\rm e}<\gamma_{\rm m}$}; \\
    \gamma_{\rm e}^{-(p+1)}, & \hbox{for $\gamma_{\rm c}<\gamma_{\rm m}<\gamma_{\rm
e}$.}
\label{eq:nakar}
\end{array}%
\right.
\end{equation}
The synchrotron radiation of electrons with a random Lorentz factor satisfying
$\gamma_{\rm c}<\gamma_{\rm e}<\gamma_{\rm m}$ and $Y(\gamma_{\rm e})\propto \gamma_{\rm e}^{-a}>1$ has a spectrum
\begin{equation}
F_\nu \propto \nu^{(a-1)/2}.
\end{equation}
As long as the IC cooling of these ``low" energy electrons is dominated by the photons in the
spectrum segment $F_\nu \propto \nu^{(a-1)/2}$, requiring that
$\Gamma_{\rm i} m_{\rm e}c^2/\gamma_{\rm e}\leq (1+z)\varepsilon_{\rm p}$, we have
\begin{equation}
Y(\gamma_{\rm e}) \propto \int^{\Gamma_{\rm i} m_{\rm e}c^2/h\gamma_{\rm e}}
{{F_\nu d\nu}\over U_{\rm B}}\propto \gamma_{\rm e}^{-(a+1)/2},
\end{equation}
where $U_{\rm B}$ is the magnetic energy density of the emitting region.
Combing with the initial assumption $Y(\gamma_{\rm e})\propto \gamma_{\rm e}^{-a}$,
we have $a=1$ and then reproduce
eqs.(22) of \citet{Deri01}
\begin{equation}
dn/d\gamma_{\rm e}\propto \gamma_{\rm e}^{-1},~~F_\nu \propto \nu^0.
\end{equation}
With eq.(\ref{eq:nakar}), it is straightforward to show that in the energy range
$\min \{1/[Y(\gamma_{\rm m})]^{2},~ [Y(\gamma_{\rm m})]^{2}\}\varepsilon_{\rm p}<h\nu<\varepsilon_{\rm p}$
\citep[see also][]{nakar09}
\begin{equation} F_\nu \propto
\left\{%
\begin{array}{ll}
    \nu^{-(p-1)/2}, & \hbox{for $Y(\gamma_{\rm m})>1$}; \\
    \nu^{-1/2}, & \hbox{for $Y(\gamma_{\rm m})<1$}.
\label{eq:nakar2}
\end{array}%
\right.
\end{equation}
Therefore only for $Y(\gamma_{\rm m})\approx 1$ the X-ray spectrum can be approximated
by $F_\nu \propto \nu^0$. $Y(\gamma_{\rm m}) \leq 1$ is also needed to satisfy the high energy radiation limit set
 by the recent Fermi observations \citep[see][and the references therein]{Fan09}.
That's why below we focus on the
ideal case $Y(\gamma_{\rm m})\approx 1$. In reality, the Band function \citep{Band93}
is smooth across $\varepsilon_{\rm p}$ and $1/2\lesssim Y(\gamma_{\rm m}) \lesssim 2$ is
allowed by the data, which helps but does not solve all the {\it fine-tuning problem}.\\

{\it The request that $Y(\gamma_{\rm m})\approx 1$ imposes a tight constraint on the magnetization of the emitting region.}
For simplicity we consider that $\beta_{\rm l}=0$ and $\beta_{\rm h}=p/2 \sim 1.25$.
The IC cooling of the electrons is dominated by the photons with an energy
$\leq \Gamma_{\rm i} m_{\rm e} c^2/\gamma_{\rm e}$. The corresponding energy density of these seed photons is $U_{\rm IC} \approx (1+z)F_0\int^{\Gamma_{\rm i} m_{\rm e} c^2/h\gamma_{\rm e}}\nu^0 d\nu \sim F_0 \Gamma_{\rm i} m_{\rm e} c^2/h\gamma_{\rm e}$.
For $\gamma_{\rm e} \sim \hat{\gamma}_{\rm m}$, the request (c) reads that
 $U_{\rm IC}\approx (\varepsilon_{\rm p}/\varepsilon_{\rm l})^{1/2}U_{\rm B}$, with which we have
\[F_0 \varepsilon_{\rm p}/h\approx F_0 \Gamma_{\rm i} m_{\rm e} c^2/h\hat{\gamma}_{\rm e,m}  \approx U_{\rm B}(\varepsilon_{\rm p}/\varepsilon_{\rm l})^{1/2},\]
where the request $\Gamma_{\rm i} m_{\rm e}c^2/\hat{\gamma}_{\rm e,m}\sim \varepsilon_{\rm p}$
 has been taken into account. The magnetic energy density $U_{\rm B} =B^2/8\pi$ and $U_\gamma$ are related by
 $U_{\rm B}=\epsilon_{\rm B}(1+\bar{Y}) U_\gamma/\epsilon_{\rm e}$.
The energy density of the soft gamma-ray emission can also be estimated by
$U_\gamma \sim {[p/(p-2)]F_0 \varepsilon_{\rm p}/h}$.
Combining all these relations we have
\[
{\epsilon_{\rm e}\over \epsilon_{\rm B}} \approx {p \over p-2}({\varepsilon_{\rm p} \over \varepsilon_{\rm l}})^{1\over 2}(1+\bar{Y}).
\]
Usually $\varepsilon_{\rm l}$ is below the low energy threshold of the detector $\sim 10$ keV. The model thus demands
\begin{equation}
{\epsilon_{\rm e}\over \epsilon_{\rm B}} \approx 100 [{p \over 5(p-2)}]({\varepsilon_{\rm p} \over 100\varepsilon_{\rm l}})^{1\over 2}({1+\bar{Y}\over 2}).
\label{eq:ratio}
\end{equation}
{\it This is a very general argument. If the low energy spectrum
$F_\nu \propto \nu^0$ is indeed produced in the way of \citet{Deri01}, we can estimate
$\epsilon_{\rm e}/\epsilon_{\rm B}$ (equally $\epsilon_{\rm e}/k$) reliably without the need of any other information expect the prompt
emission spectrum.}

With a reasonable $\epsilon_{\rm e} \leq 0.3$, eq.(\ref{eq:ratio}) gives that
$\epsilon_{\rm B} \sim 10^{-3}$, which is reasonable in the baryonic outflow model \citep[see][for a recent simulation]{ZhangW09} {\it but not
for the magnetic fireball model}.\\



{\it For completeness}, below we present the other prediction of DKK's model, which can be tested
observationally in the future. Since we find out that DKK's scenario may not apply to the magnetic fireball, here
we turn back to the standard internal shock model. With eq.(\ref{eq:gamma_em1}), we have
\begin{eqnarray}
g \sim  9~[{(1+z)\varepsilon_{\rm p}\over 600~{\rm keV}}]^{3\over
2}({\epsilon_{\rm e}\over \epsilon_{\rm B}})^{1\over
4}[(1+\bar{Y})L_{\gamma,52}]^{-1\over 4}R_{\gamma,14}^{1\over
2}\Gamma_{\rm i, 2.5}^{-1}.
\label{eq:g1}
\end{eqnarray}


Substituting eq.(\ref{eq:ratio}) into eq.(\ref{eq:gamma_em1}) and eqs.(\ref{eq:g1}),
we have
\begin{equation}
\gamma_{\rm m}\sim 5000~[{(1+z)\varepsilon_{\rm p} \over 600~{\rm keV}}]^{1\over 2}[{p\over 5(p-2)}]^{1\over 4}
({\varepsilon_{\rm p}
\over 100 \varepsilon_{\rm l}})^{1\over 8}L_{\gamma,52}^{-{1\over 4}}R_{\gamma,14}^{1\over 2},
\label{eq:gamma_em2}
\end{equation}
\begin{eqnarray}g
\sim 22~[{(1+z)\varepsilon_{\rm p}\over 600~{\rm keV}}]^{3\over
2}[{p\over 5(p-2)}]^{1\over 4}
({\varepsilon_{\rm p}
\over 100 \varepsilon_{\rm l}})^{1\over 8}L_{\gamma,52}^{-{1\over 4}}R_{\gamma,14}^{1\over 2}\Gamma_{\rm i, 2.5}^{-1}.
\label{eq:g2}
\end{eqnarray}
One can see that for typical GRB parameters,
the IC cooling of the electrons with energy $\sim \gamma_{\rm m} m_{\rm e}c^2$
is indeed within Klein-Nishina regime (i.e., $g\gg 1$). On the other hand, the request
that $\varepsilon_{\rm p}/\varepsilon_{\rm l}> g
\geq (\varepsilon_{\rm p}/\varepsilon_{\rm l})^{1/2}$ should be satisfied
otherwise the single power spectrum $F_\nu \propto \nu^{0}$ for
$\varepsilon_{\rm l}< h\nu< \varepsilon_{\rm p}$ can not hold.
With eq.(\ref{eq:g2}), such a request is equivalent to
\begin{equation}
\varepsilon_{\rm p} \gtrsim {130 \over 1+z} {~\rm keV}~({\varepsilon_{\rm p} \over 100\varepsilon_{\rm l}})^{-1/4}
[{p\over 5(p-2)}]^{-1/6}L_{\gamma,52}^{1/6}({\delta t\over 0.05~{\rm s}})^{-1/3}.
\label{eq:E_pxrf}
\end{equation}
Therefore, if DKK's scenario is the solution of the low energy spectrum problem of bright GRBs,
the X-ray flashes and X-ray flares should have a spectrum $F_\nu \propto \nu^{-1/2}$
rather than $F_\nu \propto \nu^0$, which can be tested directly. The problem is that for such soft events, the
low energy spectrum usually can not be reliably measured \citep{saka05}.
\end{appendix}


\begin{thebibliography}{}
\bibitem[Abdo et al.(2009)]{Abdo09} Abdo A. A., et al., 2009, Science,
323, 1688
\bibitem[Akerlof et al. (1999)]{Akerlof99} Akerlof C., et al. 1999, Nature, 398, 400
\bibitem[Band et al. (1993)]{Band93} Band D., et al., 1993, ApJ, 413, 281
\bibitem[Cohen et al. (1997)]{Cohen97} Cohen E., Katz J. I., Piran T., Sari R., Preece R. D., Band D. L.
1997, ApJ, 488, 330
\bibitem[Daigne et al. (2009)]{Daigne09} Daigne F., et al., 2009, in preparation
\bibitem[Daigne \& Mochkovitch (2002)]{DM02} Daigne F., Mochkovitch R., 2002, MNRAS, 336, 1271
\bibitem[Derishev (2007)]{Deri07} Derishev E. V. 2007, Ap\&SS, 309, 157 (arXiv:astro-ph/0611260)
\bibitem[Derishev et al. (2001)]{Deri01} 	Derishev E. V., Kocharovsky V. V., Kocharovsky Vl. V., 2001, A\&A, 372, 1071 (DKK)
\bibitem[Derishev et al. (2003)]{Deri03}	Derishev E. V., Kocharovsky V. V., Kocharovsky Vl. V., M\'esz\'aros P., 2003, GAMMA-RAY BURST AND AFTERGLOW ASTRONOMY 2001: A Workshop Celebrating the First Year of the HETE Mission. edited by G. R. Ricker and
    R. K. Vanderspek, AIPC, 662, 292
\bibitem[Di Matteo et al. (2002)]{Di02} Di Matteo T., Perna R., Narayan R., 2002, ApJ, 579, 706
\bibitem[Duncan \& Thompson(1992)]{dt92} Duncan R. C., Thompson C.,
1992, ApJ, 392, L9
\bibitem[Fan (2009)]{Fan09} Fan Y. Z., 2009, MNRAS, 397, 1539
\bibitem[Fan et al. (2002)]{Fan02} Fan Y. Z., Dai Z. G., Huang Y. F., Lu T., 2002,
Chin. J. Astron. Astrophys., 2, 449
\bibitem[Fan \& Piran (2006)]{FP06} Fan Y. Z., Piran T., 2006, MNRAS, 369, 197
\bibitem[Fan \& Piran (2008)]{FP08} Fan Y. Z., Piran T., 2008, Front. Phys. China., 3, 306
\bibitem[Fan et al. (2008)]{Fan08} Fan Y. Z., Piran T., Narayan R., Wei D. M., 2008, MNRAS, 384, 1483
\bibitem[Fan et al. (2004)]{Fan04}Fan Y. Z.,  Wei D. M., Wang C.
F., 2004, A\&A, 424, 477
\bibitem[Gao et al. (2009)]{Gao09} Gao W. H., Mao J. R., Xu D., Fan Y. Z., 2009, ApJ, 706, L33
\bibitem[Gendre et al. (2009)]{Gendre09} Gendre B., et al., 2009, A\&A, submitted (arXiv:0909.1167)
\bibitem[Ghisellini et al. (2000)]{ghis00} Ghisellini G., Lazzati D., Celotti A., Rees M. J., 2000, MNRAS, 316, L45
\bibitem[Giannios (2007)]{gian07} Giannios D., 2007, A\&A, 480, 305
\bibitem[Giannios \& Spruit (2006)]{gian06} Giannios D., Spruit H. C., 2006,
A\&A, 450, 887
\bibitem[Golenetskii et al. (2008)]{Gole08} Golenetskii S., Aptekar R., Mazets E., Pal'shin V., Frederiks D.,
Cline T., 2008, GCN 7482
\bibitem[Gomboc et al. (2009)]{Gomboc09} Gomboc A., et al. 2009 (arXiv:0902.1830)
\bibitem[Gotz et al. (2009)]{Gotz09} Gotz D., Laurent P., Lebrun F., Daigne F.,
Bosnjak Z., 2009, ApJ, 695, L208
\bibitem[Granot (2003)]{Granot03} Granot J.,  2003, ApJ, 596, L17
\bibitem[Greiner et al. (2009)]{Greiner09} Greiner, J., et al. 2009, A\&A, 498, 89
\bibitem[Ioka et~al.(2007)]{ioka07} Ioka K.,
Murase K., Toma K., Nagataki S., Nakamura T., 2007, ApJ, 670, L77
\bibitem[Kobayashi et al. (1997)]{KPS97} Kobayashi S., Piran T., Sari R., 1997, ApJ, 490, 92
\bibitem[K\"onigl \& Granot (2002)]{KG02} K\"onigl A., Granot J., 2002, ApJ, 574, 134
\bibitem[Kumar et al. (2007)]{kumar07} Kumar P., et al., 2007, MNRAS, 376, L57
\bibitem[Kumar \& McMahon (2008)]{km08} Kumar P., McMahon E., 2008, MNRAS, 384, 33
\bibitem[Kumar \& Panaitescu (2003)]{KP03} Kumar P., Panaitescu A., 2003,
MNRAS, 346, 905
\bibitem[Lazzati \& Begelman (2006)]{LB06} Lazzati D., Begelman M., 2006, ApJ, 641, 972
\bibitem[Levinson \& Eichler (1993)]{LE93}Levinson A., Eichler D., 1993, ApJ, 418, 386
\bibitem[Li (2008)]{Li08} Li L. X., 2008, MNRAS, 388, 1487
\bibitem[Lithwick \& Sari (2001)]{ls01} Lithwick Y., Sari R. 2001, ApJ, 555, 540
\bibitem[Lyutikov \& Blandford (2003)]{LB03} Lyutikov M., Blandford R., 2003
(astro-ph/0312374)
\bibitem[Lyutikov et al. (2003)]{Lyu03} Lyutikov M., Pariev V. I., Blandford R. D. 2003, ApJ, 597, 998
\bibitem[MacFadyen \& Woosley (1999)]{mw99}MacFadyen A. I., Woosley S. E. 1999, ApJ, 524,
262
\bibitem[M\'esz\'aros et al. (1993)]{Mesz93} M\'esz\'aros P., Laguna P., Rees M. J. 1993, ApJ, 415, 181
\bibitem[M\'esz\'aros \& Rees (2000)]{mr00} M\'esz\'aros P., Rees M. J.,
2000, ApJ, 530, 292
\bibitem[Mimica et al. (2009)]{Mimica09} Mimica P., Giannios D., Aloy M. A., 2009, A\&A, 494, 879
\bibitem[Molinari et al. (2007)]{Molinari07} Molinari E. et al., 2007, A\&A, 469, L13
\bibitem[Nakar et al. (2009)]{nakar09} Nakar E., Ando S., Sari R., 2009, ApJ in press (arXiv:0903.2557)
\bibitem[Nakar et al. (2005)]{Nakar05} Nakar E., Sari R., Piran T., 2005, ApJ, 635, 516
\bibitem[Narayan et al. (2001)]{Narayan01} Narayan R., Piran T., Kumar P., 2001, ApJ, 557, 949
\bibitem[Paczy\'nski (1990)]{Pacz90} Paczy\'nski B., 1990, ApJ, 363, 218
\bibitem[Pe'er (2008)]{peer08} Pe'er A., 2008, ApJ, 682, 463
\bibitem[Pe'er et al. (2006)]{peer06} Pe'er A., M\'esz\'aros P., Rees M.
J. 2006, ApJ, 642, 995
\bibitem[Piran (1999)]{Piran99} Piran T., 1999, Phys. Rep., 314, 575
\bibitem[Piran et al. (2009)]{psz09} Piran T., Sari R., Zou Y. C., 2009, MNRAS, 393, 1107
\bibitem[Piran et al. (1993)]{Piran93} Piran T., Shemi A., Narayan R., 1993, MNRAS, 263, 861
\bibitem[Preece et al. (2000)]{preece00} Preece R. D., Briggs M. S., Mallozzi, R. S., Pendleton G. N.,
 Paciesas W. S., Band D. L. 2000, ApJS, 126, 19
\bibitem[Rees \& M\'esz\'aros (2005)]{rm05} Rees M. J., M\'esz\'aros P., 2005, ApJ, 628,
847
\bibitem[Rossi et al. (2006)]{Rossi06} Rossi E. M., Beloborodov A. M., Rees M. J., 2006, MNRAS, 369, 1797
\bibitem[Ryde et al. (2006)]{Ryde06} Ryde F., Bj\"{o}rnsson C., Kaneko Y.,
 M\'esz\'aros P., Preece R., Battelino M., 2006, ApJ, 652, 1400
\bibitem[Sakamoto et al. (2005)]{saka05} Sakamoto T., et al., 2005, ApJ, 629, 311
\bibitem[Sari, Piran \& Narayan(1998)]{spn98}Sari R., Piran T., Narayan R., 1998, ApJ, 497, L17
\bibitem[Steele et al. (2009)]{Mund09} Steele I. A., et al. 2009, Nature, 462, 767
\bibitem[Stern \& Poutanen(2004)]{sp04}Stern B. E., Poutanen J., 2004, MNRAS, 352, L35
\bibitem[Thompson (1994)]{tho94} Thompson C., 1994, MNRAS, 270, 480
\bibitem[Toma et al. (2009)]{Toma09b} Toma K., Wu X. F., M\'esz\'aros P., 2009, ApJ
submitted (arXiv:0905.1697)
\bibitem[Usov (1992)]{usov92} Usov V. V., 1992, Nature, 357, 472
\bibitem[Wang et al. (2009)]{wang09} Wang  X. Y., Li Z., Dai Z. G., M\'esz\'aros P., 2009, ApJ, 698, L98
\bibitem[Xue et al. (2009)]{xue09} Xue R. R., Fan Y. Z., Wei D. M., 2009, A\&A, 498, 671
\bibitem[Zhang et al. (2006)]{Zhang06} Zhang B., et al. 2006, ApJ, 642, 354
\bibitem[Zhang \& Kobayashi (2005)]{ZK05} Zhang B., Kobayashi S., 2005, ApJ,
628, 315
\bibitem[Zhang et al. (2003)]{ZKM03} Zhang B., Kobayashi S., M\'esz\'aros P., 2003, ApJ,
595, 950
\bibitem[Zhang \& Pe'er (2009)]{ZhangP09} Zhang B., Pe'er A., 2009, ApJ, 700, L65
\bibitem[Zhang, MacFadyen \& Wang (2009)]{ZhangW09} Zhang W. Q.,
MacFadyen A. I., Wang P., 2009, ApJ, 692, L40
\bibitem[Zou et al. (2009)]{zfp09} Zou Y. C., Fan Y. Z., Piran T., 2009, MNRAS,
396, 1163
\bibitem[Zou \& Piran (2009)]{zp09} Zou Y. C., Piran T., 2009, MNRAS, in press (arXiv:0908.4418)
\end{thebibliography}
\end{document}